# SISTEMA OPTO-MECÁNICO PARA EL CRECIMIENTO CONTROLADO DE PILAS GRANULARES

# OPTO-MECHANICAL SYSTEM FOR THE CONTOLLED GROWTH OF GRANULAR PILES

L. Domínguez-Rubio[*], E. Martínez[*] y E. Altshuler[*]

[*]Grupo de Sistemas Complejos y Física Estadística, Facultad de Física, Universidad de La Habana, 10400 La Habana, Cuba.

Los medios granulares exhiben una notable ubicuidad en la vida humana: alimentación, industria y naturaleza son áreas donde es relevante su presencia [1-3]. También en las ciencias básicas son de utilidad, por ejemplo, al establecer analogías con eventos que ocurren en sistemas como los superconductores o el tráfico urbano [4,5].

Una configuración de granos que ha tenido mucha popularidad en la investigación y en la técnica es la pila cónica de arena. Se le puede encontrar en escenarios que van desde el almacenamiento de minerales, hasta el estudio de la criticalidad autoorganizada [1,6-9]. Mucho se ha investigado para conocer los detalles de su proceso de formación, aunque solo en configuraciones cuasi-bidimensionales se ha establecido rigurosamente una dependencia del comportamiento de la pila en crecimiento con parámetros como el flujo de entrada y la altura desde la cual se depositan los granos [9-11]. Sin embargo, los dispositivos que se han empleado para controlar experimentalmente a este último parámetro son poco precisos y dependen de un criterio de apreciación, porque se operan manualmente [10,11]. Los experimentos confirman que parámetros como el ángulo de reposo, que se puede definir como la inclinación de la superficie libre de la pila respecto a la horizontal [1,10] y que es una medida de la fricción interna de la pila, dependen de la altura de deposición de los granos [10]. Ello indica la necesidad de ejercer un control automatizado de de la altura de deposición y su influencia en las propiedades de la pila.

El objetivo de esta investigación es diseñar y evaluar un sistema que permita el control automático y preciso de la distancia entre el ápice de la pila y el punto desde donde se vierten los granos. Se empleará el procesamiento de imágenes como herramienta para cuantificar los resultados y evaluar las fuentes de incertidumbres que posee el sistema. El principio de funcionamiento del mecanismo de control de la altura es la obstrucción por la pila de una señal luminosa láser. Así se indica que la separación entre el ápice de la pila y el punto desde donde se vierten los granos es menor que el valor prefijado, y debe elevarse la posición del sistema de suministro para mantener esa altura constante.

En la figura 1 a) se diagraman las partes componentes del sistema de control: El contenedor (1) suministra el material granular (arenas de diversos tipos) a través de un orificio con diámetro variable. Al crecer la pila (2) se interrumpe el haz del láser (3) y el circuito de control (4) activa el motor (5) que hace girar la polea (6) y asciende el contenedor hasta que vuelve a llegar luz al sensor. Para asegurar que no deslice la correa en el eje del motor y aumentar la precisión del movimiento ascendente del contenedor se coloca el contrapeso (7). El crecimiento de la pila es registrado por las cámaras (8) y (9) las cuales graban, respectivamente, la base y el perfil lateral de la pila. Con este dispositivo se pueden regular los parámetros esenciales que controlan la dinámica de la formación de las pilas: flujo de suministro de granos y altura de deposición. La cámara 9 será utilizada para evaluar la efectividad del sistema de control de altura de suministro, como veremos más adelante.

El diseño del circuito de control es también muy simple: al obstruirse el haz láser que incide en el fotodiodo, este deja de conducir y la corriente va a la base del transistor; se establece entonces una corriente emisor-colector que pasa por el *relay* que cierra el circuito del motor. Cuando el motor gira, sube el contenedor hasta que el láser sobrepasa el ápice de la pila que lo de modo que vuelve a incidir en el fotodiodo. Este lleva la corriente de base a tierra y se elimina su paso por el transistor. El *relay* abre entonces el circuito del motor y el contenedor deja de subir.

Debe cuidarse que la columna de granos no obstruya completamente al haz láser. Para ello se coloca al sistema láser-sensor desplazado lateralmente del chorro como se muestra en la figura 1 b). Esto implica que no se registren las variaciones exactamente del ápice de la pila, el cual se ubica en el centro de la columna. Debido a los choques de los granos, el tope de la pila es frecuentemente un *plateau* y en su borde se puede ubicar el eje del sistema sensor-láser: se establece así un punto de referencia desplazado lateralmente respecto al ápice y, cuando no sea un *plateau* el tope, estará en una posición inferior. La distancia entre el punto de referencia y la abertura será la altura que se debe mantener constante, denominada como *altura de suministro*.

Un limitación del método es que el dispositivo solo tiene un movimiento ascendente y regula variaciones en la altura prefijada cuando el valor es menor que el establecido por la distancia abertura-punto de referencia. Esta dificultad impone que las fluctuaciones en la dirección vertical medidas respecto a tierra de este punto de referencia, deben tributar a la incertidumbre con la que se computa la altura de suministro. No es conveniente, además, que el dispositivo reproduzca los movimientos de ascenso y descenso del tope de la pila porque podrían inducirse inestabilidades en el flujo de granos.

Con los videos del crecimiento de las pilas se obtiene el valor de la altura de deposición y su rango de incertidumbre. Se escribió un programa en *C++* utilizando la biblioteca de visión por computadoras *OpenCV* para el procesamiento de los videos. El programa binariza las imágenes de las pilas y, mediante la construcción del diagrama espacio-temporal que se muestra en la figura 2 a), se puede conocer la distancia desde el punto de referencia de la pila hasta la abertura para cada cuadro del video.

Se evaluará ahora la efectividad de nuestro sistema mediante el análisis sistemático de imágenes laterales de video capaces de capturar las posiciones respecto a tierra de la abertura, del ápice de la pila y del punto de referencia sobre su superficie. A este proceso contribuyen principalmente dos fuentes de incertidumbre: (a) las fluctuaciones en la separación entre la abertura y el punto de la superficie de la pila que el láser utiliza como referencia y (b) la incertidumbre asociada al procesamiento de video, cuyo valor más importante está dado por el cómputo del umbral para convertir las imágenes a blanco y negro).

Es fácil demostrar que las incertidumbres en el proceso de binarizar las imágenes pueden no considerarse: el cambio de los umbrales de binarización en un rango amplio arrojó diferencias de alturas del orden de 2 píxeles, equivalentes a menos de 1 mm en el experimento real. Este valor es menor que las fluctuaciones en la altura del punto de referencia en la pila respecto a tierra, con un valor de aproximadamente 3 mm. El análisis de los videos nos permitió determinar la evolución temporal de la distancia entre la abertura y el punto de referencia utilizado por el haz de láser para así computar sus fluctuaciones y compararlas con el valor medio de la distancia abertura-punto de referencia.

El diagrama espacio-temporal que se ilustra en el panel inferior de la Figura 2 a) muestra, a través del ancho de la franja negra, la evolución de la distancia abertura-punto de referencia de la pila. Los gráficos de la derecha exhiben la evolución de la distancia hallada a través de los diagramas espacio-temporales para diversos flujos de entrada ($0.6\ g/s$, $1.2\ g/s$ y $11.4\ g/s$) y para diversas alturas medias. En general puede establecerse que la altura de suministro se mantiene constante tanto cualitativa como cuantitativamente. Para las configuraciones graficadas en la figura 2 b) las fluctuaciones son menores que el 10% de la distancia abertura-punto de referencia en los casos 1, 2 y 3. Sólo en el caso 4 estas fluctuaciones son comparables con la distancia media, lo que indica que el control no es confiable. Sin embargo, la situación ilustrada en 4 es completamente esperable porque se trata de una distancia abertura-punto de referencia del orden del ancho del haz láser utilizado. En resumen, nuestro sistema, para la mayoría de las situaciones de interés experimental, puede controlar confiablemente la distancia en cuestión, por debajo de un 10% de incertidumbre relativa. Además, al determinar la evolución temporal de las trayectorias verticales de la abertura y del punto de referencia se puede comprobar que ambas coinciden y esto sirve como legitimación adicional del buen funcionamiento del dispositivo.

Un último elemento a analizar en la Figura 2, es la comparación entre las fluctuaciones de la posición de la abertura y las del punto de referencia sobre la superficie de la pila medida respecto a tierra. Esto nos permitirá determinar cuál factor es el que tributa más significativamente a la incertidumbre de la altura de suministro, si las deficiencias en el sistema electromecánico o las fluctuaciones del tope de la pila producto al impacto de los granos. Con el diagrama espacio-temporal se puede obtener que, aunque las fluctuaciones respecto a tierra del punto de referencia y de la abertura son del mismo orden de magnitud, el "ruido" que se observa en los gráficos 1-4 de la figura 2 b) es de "alta frecuencia", y se corresponde con la frecuencia de las fluctuaciones del punto de referencia y no con las de la abertura. El análisis de conjunto de la figura 2 también confirma que nuestro sistema no es capaz de "seguir" esas fluctuaciones en detalle, pero esto sería un inconveniente para nuestros objetivos experimentales, como se apuntó anteriormente.

Una evaluación adicional es determinar el coeficiente de restitución ($e$) del material granular. Su cómputo usual [1] se realiza mediante la expresión $e^2 \sim 1 - (\Delta E_c / E_c)$, donde $E_c$ es la energía cinética antes del choque y $\Delta E_c$, la variación luego de ocurrir. Un buen estimado de esa razón entre energías, en nuestro caso, es $\Delta h / h$ donde $h$ es la altura de suministro que se supone constante y $\Delta h$, la incertidumbre de esta altura dada por la desviación estándar de los valores medidos, que es una medida de los cambios en la altura del nuevo ápice a causa de los impactos. El valor promedio que se obtiene por este método para $\Delta h / h$ es de 0.11, lo que establece un coeficiente de restitución próximo a 0.9, en acuerdo con lo que se informa en la literatura para el material granular empleado [1].

Probar la efectividad de este sistema electromecánico para el control preciso de la altura de suministro de granos permitirá evaluar, de manera controlada, la influencia de los cambios de esta magnitud en las características y la dinámica de crecimiento de las pilas. Su aplicabilidad puede extenderse a configuraciones cuasi-bidimensionales, donde también el dispositivo opera con éxito según las experiencias más recientes de nuestro grupo de investigación.

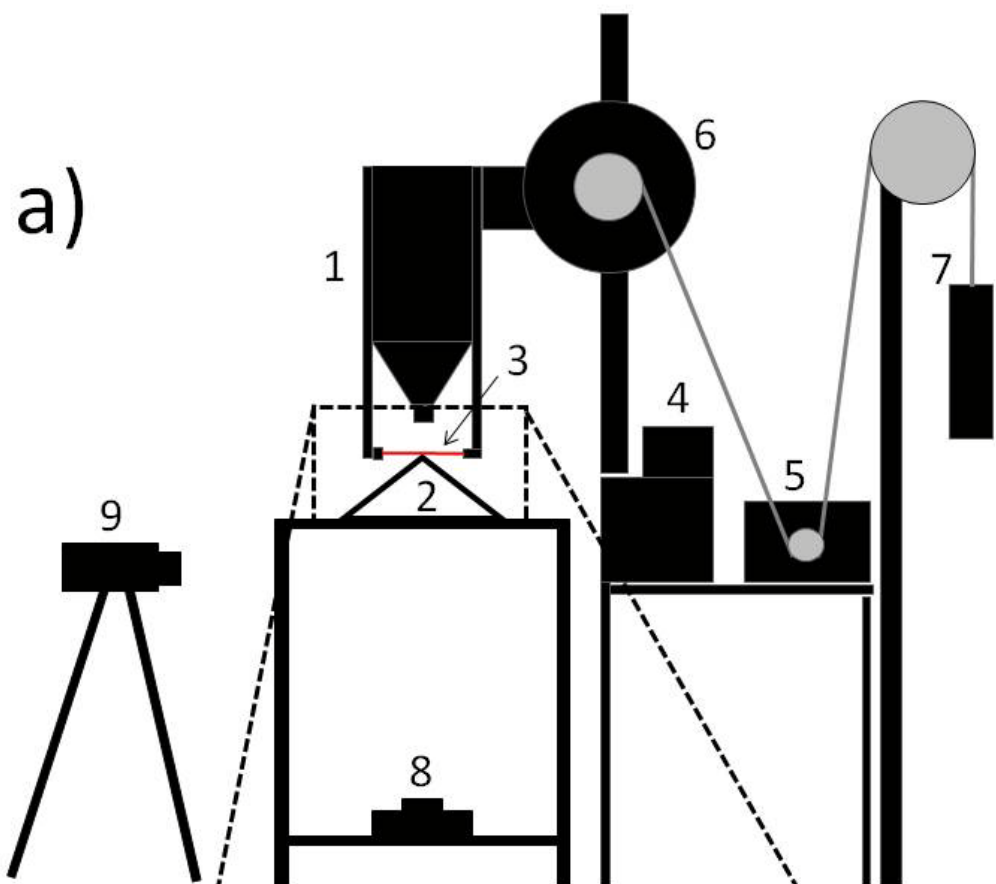

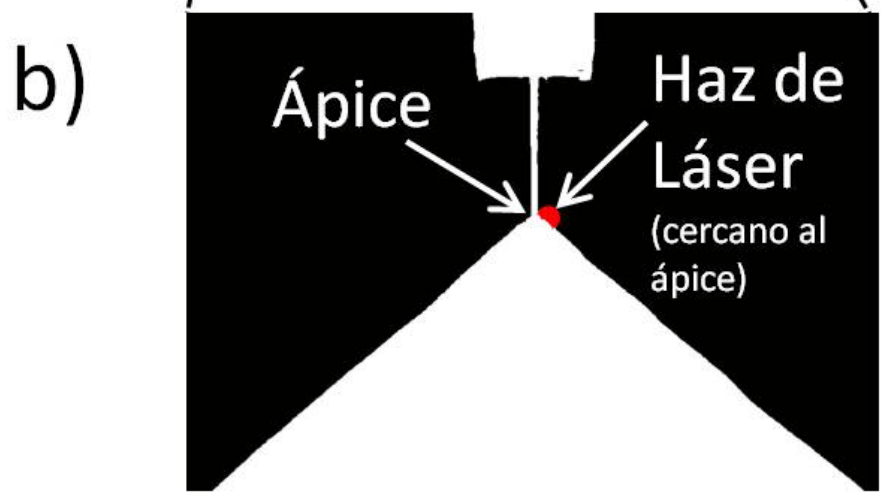

**Figura 1**. Esquema del dispositivo experimental. a) Diagrama de las partes componentes del sistema de control: (1) contenedor con apertura de diámetro variable, (2) pila de granos, (3) haz láser, (4) circuito de control, (5) motor, (6) polea, (7) contrapeso , (8) y (9) cámaras. b) Vista lateral de la pila de granos donde se muestra el haz láser desplazado lateralmente respecto a la columna de granos sobre una región del borde de la pila que sirve como punto de referencia. El ápice puede estar aplanado a causa del impacto de los granos.

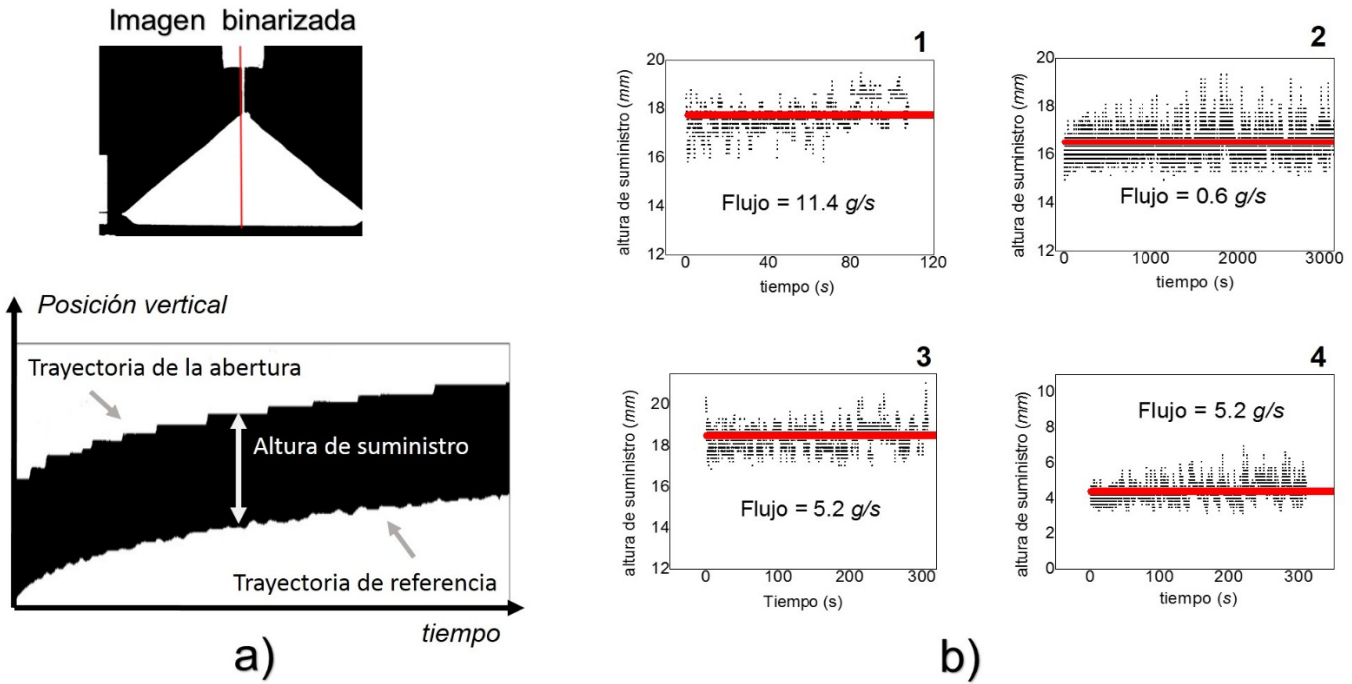

**Figura 2**. Procesamiento de imágenes y sus resultados: a) En el panel superior se muestra un cuadro binarizado del video que registró el crecimiento de la pila. Para la construcción del diagrama espacio temporal del panel inferior se toma como referencia la línea vertical que aparece sobre la imagen de la pila. b) Valores medidos de la altura de suministro y el ajuste teórico a una constante para diferentes flujos de entrada de granos